\documentclass[english,aps,prper,reprint,showpacs,titlepage,longbibliography]{revtex4-2}   
\usepackage[utf8]{inputenc}
\usepackage[compatibility=false]{caption} 
\usepackage{booktabs} 
\usepackage{tabularx} 
\usepackage{ragged2e} 
\usepackage{caption} 
\usepackage[T1]{fontenc}	
\usepackage{geometry}
\usepackage{times}
\usepackage{hyperref}  
\hypersetup{colorlinks=true,urlcolor=blue,citecolor=blue,linkcolor=blue} 
\usepackage{array}
\usepackage{enumerate}
\usepackage{amsmath}
\usepackage{amssymb}
\usepackage{tikz}
\usepackage{graphicx}
\usepackage{multirow}
\usepackage{tcolorbox}
\usepackage{ragged2e}
\usepackage{float}
\usepackage[utf8]{inputenc}

\usepackage[normalem]{ulem}
\usepackage{setspace} 
\usepackage{natbib}
\begin{document}

\title{Probing AI-generated physics solutions and preparing students to critique them}

\author{Nikhil Sanjay Borse}
 \affiliation{Department of Physics and Astronomy, Purdue University, 525 Northwestern Ave, West Lafayette, IN-47907, U.S.A.}

\author{Amir Bralin}
\affiliation{Department of Physics and Astronomy, Texas Tech University, 2500 Broadway, Lubbock, TX-79409, U.S.A.} 

\author{Sean Savage}
\affiliation{Department of Physics, Florida International University, 11200 SW 8th St, Miami, FL-33199, U.S.A.} 
  
\author{N. Sanjay Rebello}
\affiliation{Dept. of Physics and Astronomy / Dept. of Curriculum \& Instruction, Purdue University, West Lafayette, IN-47907, U.S.A.} 

\keywords{}

\begin{abstract}

This study examines Artificial Intelligence (AI)-generated physics solutions from two connected perspectives: how prompt design shapes these solutions and how students can be prepared to critique them. Using a rotational-mechanics problem, we adapted a problem-classification framework to examine prompt variations, evaluating OpenAI's o4-mini responses with the Minnesota Assessment of Problem Solving (MAPS) rubric. Well-specified prompts improved solution completeness; underspecified and multimodal prompts exposed weaknesses in physics reasoning and correctness. In the student-evaluation phase, 24 introductory physics lab groups evaluated an o4-mini solution to this problem after either independently solving a related problem or critiquing its AI-generated solution with MAPS-based reflection questions. Problem-solving-only groups exhibited uncritical or misconception-based critiques; MAPS-guided groups identified more expert-aligned issues, including skipped numerical procedures and undefined notation. Together, our findings contribute to physics education research by showing how AI-generated solutions can ground both model-reasoning benchmarks and improved student critique of that reasoning through MAPS-based reflection.

\clearpage
\end{abstract}

\maketitle

\section{Introduction \&\ Background}

Large Reasoning Models (LRMs) such as OpenAI's o4-mini show promise in solving physics problems requiring multi-step reasoning and qualitative interpretation \cite{OpenAIOA,horchani2025chatgpt}. However, prior work has largely benchmarked LLMs across various problem types using fixed, fully specified prompts, while less is known about how modifying the same problem statement affects model reasoning. This study addresses that gap through a focused investigation of one novel rotational-mechanics problem family involving qualitative reasoning and graphical prediction. Grounded in Jonassen’s problem classification framework, we adapt two dimensions for generative-AI problem solving: \textit{prompt-specificity}, or how much information is explicitly given versus left for the model to infer, and \textit{modality}, or whether the problem is presented through text alone or with visual representations \cite{jonassen2000toward}. By progressively removing details from a well-specified prompt and varying text–visual combinations, we examine how LRMs reason when constraints must be inferred or interpreted across representations. This single-problem deep dive parallels Dunlap et al.'s focused study of inclined-plane problem variations \cite{PhysRevPhysEducRes.21.010153}, but extends that approach by systematically controlling prompt-specificity and modality with a reasoning-capable LRM. The design also connects to problem-posing traditions in education and to physics education research (PER) work on educational data augmentation and synthetic assessment generation \cite{english1997development,silver1994mathematical,stoyanova1996framework,singer2013problem,kieser2023educational}. We evaluate model output using not only problem-solving accuracy \cite{wang2024examining}, but also the Minnesota Assessment of
Problem Solving (MAPS) rubric, which assesses \textit{useful descriptions, physics approach, specific application, mathematical procedures}, and \textit{logical progression} \cite{docktor2016assessing}. Because the task requires predicting and graphing height evolution, it builds on research documenting student difficulties with motion representation and graph interpretation \cite{mcdermott1987student,beichner1994testing,van1991learning}.

The same features that make AI-generated physics solutions useful for instruction also make them important objects of critique. Generative AI can produce fluent, plausible, and partially correct responses that still contain conceptual errors, unsupported assumptions, incomplete reasoning, or misleading representations. Kortemeyer showed that ChatGPT could approach passing an introductory physics course while still exhibiting pedagogically significant errors; Bralin et al. found that Open AI's o3-mini model follows the verbal reasoning it generates, while solving physics problems, but does not evaluate its intermediate steps; Polverini et al. found that multimodal AI systems can struggle with visual interpretation and physics representations; and Shojaee et al.'s study suggested that LRMs can display systematic, complexity-dependent reasoning limitations \cite{PhysRevPhysEducRes.19.010132,6fmx-bsnl,bralin2025,casal2023ai,Polverini_2025,shojaee2025illusion}. If such limitations are partly predictable, then students must learn to identify recurring weaknesses, including mismatches with the problem statement, unjustified physics approaches, skipped mathematical steps, unreliable diagrams or plots, and reasoning that appears coherent without being fully justified.

Despite growing research on AI performance in physics, less is known about how students can be prepared to critique AI-generated physics solutions. Dahlkemper et al. found that students' evaluations of ChatGPT responses depended on scientific accuracy, linguistic quality, and self-assessed knowledge; Ding et al. highlighted perceived accuracy in students’ interactions with ChatGPT as a physics tutor; and Bitzenbauer proposed using ChatGPT outputs to foster critical thinking, though such work remains exploratory \cite{Dahlkemper_2023,ding2023students,bitzenbauer2023chatgpt}. \par Building on this prior work, this study investigates strategies to facilitate students to critically evaluate AI-generated solutions of physics problems. Specifically, we compare two conditions for preparing students to critique AI-generated solutions to a physics problem: 1) Solving a related physics problem independently, and 2) critiquing an AI-generated solution to the same related problem using reflection prompts based on the MAPS rubric \cite{docktor2016assessing}. This comparison builds on prior work contrasting problem solving with example-based reflection, while leaving open which form of preparation better supports students’ critique of AI-generated reasoning \cite{chi1989self,booth2013using}. Both conditions then evaluate an AI-generated solution to a related, but more complex problem. This design responds to emerging STEM assessment work in which students compare their reasoning with AI-generated output, while extending it by examining transfer to a more complex task \cite{de2026stem}. Research on transfer and preparation for future learning further motivates this design because the common test problem asks whether students can apply critique practices beyond the initial preparation task \cite{bransford1999chapter,nokes2013toward}.

In sum, this paper examines LRM physics reasoning both as a model output shaped by prompt variations as well as a resource for learning problem-solving that students must learn to critique. The first part uses Jonassen-informed prompt variation and MAPS-based analysis to identify weaknesses in an LRM's solution to a rotational-mechanics problem. The second part asks whether conventional problem-solving or structured MAPS-based critique better prepares students to evaluate the same LRM solution. Together, these aspects support a broader PER goal: characterizing recurring weaknesses in AI-generated physics reasoning and helping students develop calibrated, evidence-based critiques of AI-generated problem solutions. Our research questions are:

\textbf{RQ1.} How does varying the prompt-specificity and modality of a single physics-problem prompt affect accuracy and MAPS rubric scores for an LRM such as o4-mini? 

\textbf{RQ2.} How do different strategies for preparing students to critique AI-generated solutions shape the quality of students' MAPS-aligned critiques, justifications, and perceived-accuracy judgments of an AI-generated solution to a related but more complex physics problem?

\begin{figure*}[t]
    \centering
    \setlength{\tabcolsep}{1pt}

    \begin{tabular}{cc}
        \includegraphics[
            width=0.22\textwidth,
            trim=125 70 150 50,
            clip
        ]{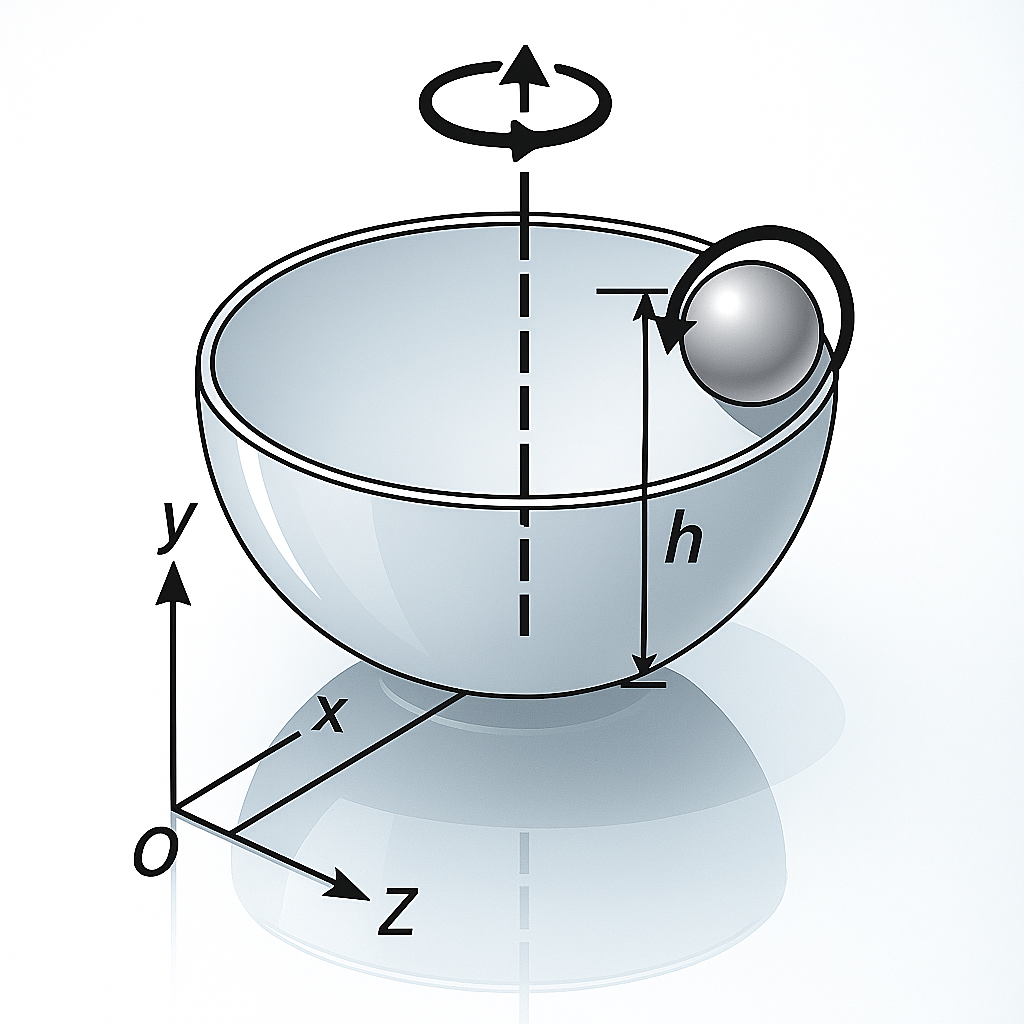} &
        \includegraphics[width=0.77\textwidth]{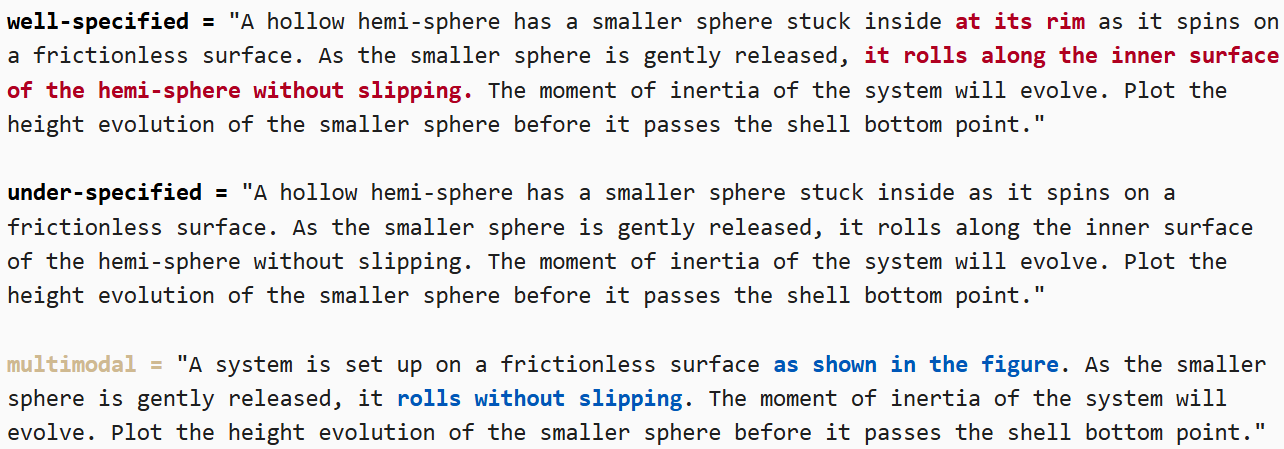}
    \end{tabular}

    \caption{Three prompt variations used in the study: text-only prompts are well-specified and underspecified, respectively, while the multimodal prompt includes the AI-generated diagram (left); bold colored text marks key differences.}
    \label{fig:o4-mini-prompt-variants}
\end{figure*}

\section{Methods}

We adapt Jonassen's framework to LRM reasoning by varying prompts along two dimensions: prompt-specificity and modality \cite{jonassen2000toward}. For text-only prompts, we systematically remove details from a well-specified prompt so that the LRM must make additional assumptions; for modality variations, we augment the same text with diagrams. The task is a rotational-mechanics problem involving a smaller sphere rolling without slipping along the inner surface of a hollow rotating hemisphere. Although the broader analysis included additional prompt variants, this paper focuses on three representative cases: a well-specified text-only prompt, an underspecified text-only prompt, and a multimodal prompt combining text with an AI-generated diagram chosen for clarity. Figure \ref{fig:o4-mini-prompt-variants} shows these variants. The model under study is o4-mini, selected for its multi-step reasoning capabilities \cite{OpenAIOA}. Data were collected through the OpenAI API. Each prompt variant was submitted five times using default settings, and complete outputs, including stepwise reasoning text were collected \cite{OpenAIOA}.

\begin{table}[t]
\centering
\caption{MAPS rubric scores for the three prompt variants.}
\label{tab:maps-rubric-scores}
\renewcommand{\arraystretch}{1}
\setlength{\tabcolsep}{3pt}
\scriptsize

\resizebox{\columnwidth}{!}{%
\begin{tabular}{@{}lccc@{}}
\hline
MAPS criterion &
\shortstack{well-specified} &
\shortstack{underspecified} &
\shortstack{multimodal} \\
\hline
Useful Descriptions & 5 & 5 & 4 \\
Physics Approach & 5 & 4 & 4 \\
Specific Application & 5 & 5 & 5 \\
Math Procedures & 4 & 3 & 4 \\
Logical Progression & 5 & 3 & 4 \\
\hline
\end{tabular}%
}
\end{table}

In Phase I, AI outputs to the three prompt variants were evaluated using two measures: accuracy, defined as agreement with the correct quantitative result under explicit or implicit assumptions, and MAPS rubric scores \cite{docktor2016assessing}. MAPS evaluates five dimensions of problem solving on a five-point Likert scale: \textit{useful descriptions} (i.e., relevant representations and given/target quantities); \textit{physics approach} (i.e., appropriate principles); \textit{specific application} (i.e., correct contextual use of principles); \textit{mathematical procedures} (i.e., valid algebraic, computational, and graphical work); and \textit{logical progression} (i.e., coherent reasoning) \cite{docktor2016assessing}. To establish reliability for MAPS scoring of the more involved common solution later critiqued by students, two expert raters independently scored the solution, and agreement was measured using weighted Cohen's $\kappa$ with linear weights \cite{tinsley1975interrater,de2021comparison}.

In Phase II we examined how preparation conditions affected students' critiques of AI outputs. The activity was conducted in a calculus-based, first-semester undergraduate physics course taken primarily by future engineering majors at a large U.S. Midwestern university. All data were anonymized before analysis so that participant identities were not available to the researchers. Data came from 24 student groups of three students, drawn from five laboratory sections and representing a subset of 1,759 students (587 groups) enrolled in the course. In the `Solution' Condition, 15 groups across three sections solved a textbook-style rolling-sphere inclined-plane problem and qualitatively graphed height evolution without AI. In the `Critique' Condition, 9 groups across two sections critiqued an AI output to the same preparation problem using 15 reflection questions organized around five MAPS categories, three per category, rather than the full MAPS rubric or scoring-level descriptors. The time on task for both conditions was approximately 30 minutes.

After preparation, both conditions completed a common task in which they evaluated the multimodal LRM solution to the rotating-bowl problem. Students rated the solution through five scoring questions aligned with MAPS categories and justified each score in writing. They also answered two follow-up survey questions about perceived accuracy of the specific AI output and of AI-generated solutions in general. Student data-handwritten solutions from the `Solution' Condition, reflection responses from the `Critique' Condition, MAPS-aligned ratings, justifications, and survey responses were analyzed qualitatively to identify patterns in students' critiques, score justifications, and perceived accuracy \cite{miles2014qualitative}.

\section{Findings \&\ Discussion}

\subsection{Phase I: Prompt variation and model reasoning}

For the well-specified prompt, o4-mini produced a mostly complete analytical and numerical solution, including the expected quantitative height evolution. Its main weakness was an incomplete explanation of some mathematical procedures, particularly the transition from the physical setup to the analytical formulation used for computation. For the slightly underspecified prompt, o4-mini identified much of the relevant physical structure, but its numerical implementation failed. For the multimodal prompt, o4-mini recovered much of the analytical structure but skipped the numerical integration required for a quantitative result, instead producing only a qualitative final response and verbal description.

Table \ref{tab:maps-rubric-scores} shows the MAPS rubric scores for the solution generated in response to each prompt. Across pure text-only prompts, only the well-specified prompt produced a correct quantitative result. The MAPS scores, however, revealed differences in reasoning quality beyond final accuracy. The well-specified prompt scored 5 in all MAPS categories except \textit{Mathematical Procedures}, where it received a 4 because some steps were not fully explained. The underspecified prompt received a 4 in \textit{Physics Approach} because it neglected aspects of the hemisphere's rotational kinetic energy and angular-momentum conservation, and 3 in \textit{Mathematical Procedures} and \textit{Logical Progression} because its numerical errors produced a final result inconsistent with the rest of the solution. The multimodal prompt scored 4 in \textit{Useful Description}, \textit{Physics Approach}, \textit{Mathematical Procedures}, and \textit{Logical Progression} because it lacked supporting representations such as a free-body diagram, did not define the total kinetic energy or variables like rolling angle $\Psi$ clearly, and produced only a qualitative rather than quantitative final result.

These patterns suggest that prompt-specificity and modality shaped both correctness and reasoning quality. Across the five trials for each prompt, the AI outputs varied somewhat in organization and presentation, but were generally consistent in their final outcomes. The well-specified prompt consistently elicited more complete solutions, whereas the underspecified prompt trials typically showed numerical issues that resulted in incorrect or otherwise misleading final results. The multimodal prompt trials consistently skipped the numerical integration needed for a quantitative height-evolution result, instead producing qualitative final responses and verbal descriptions. These findings suggest that underspecification increased the likelihood of missing assumptions, oversimplified physics, and numerical or representational errors. This result aligns with prior PER work on the role of problem structure in eliciting reasoning and with studies showing that prompt structure can affect AI-generated solutions \cite{jonassen2000toward,mcdermott1999resource,PhysRevPhysEducRes.21.010153,kieser2023educational}. The multimodal result also reinforces concerns that AI systems may struggle with visual representations even when relevant supporting text is provided \cite{Polverini_2025}. Two expert raters independently scored the AI outputs using MAPS and showed substantial agreement, with weighted Cohen's $\kappa = 0.64$ \cite{de2021comparison,tinsley1975interrater}. Other disagreements were limited to how strongly to penalize the absence of a free-body diagram for the multimodal prompt and the treatment of rolling-angle derivative for the underspecified prompt, thus supporting MAPS as a useful lens for distinguishing correctness, physics reasoning, mathematical execution, and coherence.

\subsection{Phase II: Student preparation and critique quality}

We examined whether different preparation conditions helped students notice comparable issues in the o4-mini solution of a related problem. Preparation-phase data showed that groups in both student conditions generally completed the intended activity. In the `Solution' condition, most groups correctly derived the height-evolution equation for a sphere rolling down an inclined plane, although some did not produce the requested qualitative plot. In the `Critique' condition, most groups completed nearly all of the MAPS-based reflection questions. When asked what part of the solution they would verify before accepting it, some groups identified the qualitative plot for numerical verification, echoing expert critiques in Phase I of the study. Thus, both preparation conditions produced sufficient engagement for comparing how they shaped students' critiques of a common LRM solution to the multimodal prompt in Phase I. Given group-level artifacts, we interpret the student component as an exploratory comparison of critique patterns.

The `Solution' condition responses showed three qualitative patterns. Many groups accepted the AI output largely because it appeared polished and organized, assigning high MAPS scores without clear evidence of checking the problem statement, physics, or graphical output. This pattern is consistent with prior work showing that students' judgments of AI responses can be influenced by linguistic quality or apparent coherence \cite{Dahlkemper_2023,ding2023students}. Several groups were skeptical but critiqued the solution for reasons that often reflected their own traditional misconceptions rather than expert-aligned concerns. Other groups offered more targeted critiques, noting missing diagrams, skipped steps, insufficient explanation, or weak representational support, which aligns with the intent of the MAPS rubric \cite{docktor2016assessing}.

Responses in the `Critique' condition showed a stronger focus on features of the AI-generated solution as an object of critique. Several groups identified the mismatch between the AI solution's qualitative plot and the numerical integration needed for a quantitative height-evolution graph, closely matching the expert critique of the multimodal response. Some groups also noted confusion in the polar-angle limits. Others focused on alignment with course conventions, noting that the solution used rolling angle $\Psi$ without defining it clearly and introduced dot notation only after it had already appeared. These critiques show that students were evaluating not only the correctness of the AI-generated solution, but also whether it was understandable and aligned with the problem-solving strategies used in their course.

Students' perceived-accuracy responses reinforced the patterns described above. In the `Solution' condition, groups that accepted the solution uncritically tended to report relatively high confidence in the specific AI solution while remaining cautious about AI tools in general. Groups whose critiques reflected misconceptions reported low perceived accuracy, but their skepticism was not always well grounded. Groups that identified missing diagrams, skipped steps, or weak explanations expressed more calibrated trust, viewing AI as useful only when carefully checked. In the `Critique' condition, confidence was also moderate, but skepticism was more often tied to valid concerns: the lack of numerical integration, unclear notation, qualitative rather than quantitative plotting, and misalignment with course methods. One group explicitly noted that AI models struggle with problems involving visual representations, echoing both the expert critique and prior findings on multimodal AI limitations in physics \cite{Polverini_2025}.

Overall, both conditions supported productive critique, but in different ways. The `Solution' condition facilitated students to engage with the underlying rolling-motion physics, yet a substantial fraction either accepted the AI solution too readily or critiqued it based on their own misunderstandings. `Critique' condition responses showed more frequent attention to the kinds of AI-solution weaknesses identified in the model analysis, including missing definitions, unclear notation, skipped computational steps, and confusion between qualitative and quantitative representations. These exploratory results suggest that MAPS-based reflection may support more expert-aligned critique than independent problem solving alone. This finding shows that evaluating AI-generated work requires not only physics knowledge, but also practice applying disciplinary criteria to the reasoning, representation, and communication of a solution \cite{docktor2016assessing,bitzenbauer2023chatgpt,Dahlkemper_2023}.

\section{Conclusion, Limitations, \&\ Future Work}

This study examined how AI-generated physics reasoning changes across prompt conditions and how students can be prepared to critique the resulting strengths and weaknesses. o4-mini performed strongly on the well-specified prompt, deriving an appropriate analytical setup and producing a quantitative height-evolution result, but its performance weakened when prompt details or representations changed. Underspecified prompts led to missing assumptions, a weaker physics approach, and misleading quantitative results despite a plausible analytical setup, while the multimodal prompt produced a qualitatively reasonable but mathematically incomplete response by skipping numerical integration. These findings reinforce the need to evaluate AI-outputs not only by final answers, but also by their physics approach, mathematical procedures, representations, and logical coherence. Together, Jonassen's framework and the MAPS rubric provide a structured way to probe these dimensions of LRM reasoning. Our findings are thus consistent with recent work suggesting that reasoning-model errors may be partly predictable as problem complexity changes, including failures in exact computation and inconsistent reasoning across problem scales \cite{shojaee2025illusion}.

The student component extends these implications to instruction. Most groups completed their preparation tasks successfully, but the test phase showed important differences in critique quality. Some groups that first solved a related problem accepted the AI solution too readily, while others were skeptical for reasons that reflected their own misconceptions. In contrast, several groups that completed MAPS-guided reflection identified issues more closely aligned with expert critiques, including missing numerical integration, qualitative rather than quantitative plotting, unclear notation, and misalignment with course conventions. This pattern was consistent with their preparation-phase reflections, where some Critique-condition groups had already identified graphical output as something they would want to verify before accepting an AI-generated solution. Perceived-accuracy responses showed a similar pattern: students were generally cautious about AI, but MAPS-guided groups often grounded their caution in specific, defensible features of the solution.

Overall, the findings suggest that conventional problem solving and AI-solution critique are related but distinct forms of preparation. Solving a related problem can support physics knowledge, but it alone may not prepare students to evaluate the distinctive strengths and weaknesses of LRM reasoning. MAPS-guided critique may help students treat AI output as an object of disciplinary evaluation, attending to whether representations match the problem, variables are defined, procedures are completed, and reasoning is useful for learners. As AI tools improve, their outputs may become more fluent and difficult to evaluate through surface-level cues alone, making structured critique increasingly important for identifying subtle inconsistencies, missing assumptions, and incomplete reasoning. Our broader contribution is thus an integrated PER approach: using prompt variation to identify recurring weaknesses in AI outputs, and using those weaknesses to design activities that help students evaluate AI reasoning with disciplinary criteria rather than surface-level trust.

This study is limited by its use of one model, o4-mini, and one rotational-mechanics problem family, so findings may not generalize across models, topics, or domains; however, the method of varying prompt-specificity and modality with MAPS-based evaluation remains transferable. Student analysis was limited to 24 groups from five lab sections, with conditions assigned by section and group artifacts analyzed rather than individual reasoning. Given these design constraints, the student component should be viewed as an exploratory comparison of critique patterns. Future work should use larger samples across instructors, include individual reflections or interviews, and compare multiple LRMs and physics topics. Finally, because students critiqued a fixed AI-generated solution, future studies should examine more authentic AI use, including students' own prompts, follow-up questions, verification practices, and transfer of critique skills across text-only, diagram-based, and computational AI responses.

\section{Acknowledgments}

This work is supported in part by U.S. National Science Foundation Grant 2111138. Any opinions expressed are those of the authors and not the Foundation.

\clearpage
\bibliography{references}

@article{docktor2016assessing,
  title={Assessing student written problem solutions: A problem-solving rubric with application to introductory physics},
  author={Docktor, Jennifer L and Dornfeld, Jay and Frodermann, Evan and Heller, Kenneth and Hsu, Leonardo and Jackson, Koblar Alan and Mason, Andrew and Ryan, Qing X and Yang, Jie},
  journal={Physical review physics education research},
  volume={12},
  number={1},
  pages={010130},
  year={2016},
  publisher={APS}
}

@article{tinsley1975interrater,
  title={Interrater reliability and agreement of subjective judgments.},
  author={Tinsley, Howard E and Weiss, David J},
  journal={Journal of Counseling Psychology},
  volume={22},
  number={4},
  pages={358},
  year={1975},
  publisher={American Psychological Association}
}

@article{PhysRevPhysEducRes.21.010153,
  title = {Descending an inclined plane with a large language model},
  author = {Dunlap, Justin C. and Sissons, Ryan and Widenhorn, Ralf},
  journal = {Phys. Rev. Phys. Educ. Res.},
  volume = {21},
  issue = {1},
  pages = {010153},
  numpages = {18},
  year = {2025},
  month = {May},
  publisher = {American Physical Society},
  doi = {10.1103/PhysRevPhysEducRes.21.010153},
  url = {https://link.aps.org/doi/10.1103/PhysRevPhysEducRes.21.010153}
}

@article{jonassen2000toward,
  title={Toward a design theory of problem solving},
  author={Jonassen, David H},
  journal={Educational technology research and development},
  volume={48},
  number={4},
  pages={63--85},
  year={2000},
  publisher={Springer}
}

@article{kieser2023educational,
  title={Educational data augmentation in physics education research using ChatGPT},
  author={Kieser, Fabian and Wulff, Peter and Kuhn, Jochen and K{\"u}chemann, Stefan},
  journal={Physical Review Physics Education Research},
  volume={19},
  number={2},
  pages={020150},
  year={2023},
  publisher={APS}
}

@article{mcdermott1999resource,
  title={Resource letter: PER-1: Physics education research},
  author={McDermott, Lillian C and Redish, Edward F},
  journal={American journal of physics},
  volume={67},
  number={9},
  pages={755--767},
  year={1999},
  publisher={American Association of Physics Teachers}
}

@inproceedings{OpenAIOA,
  title={OpenAI o3 and o4-mini System Card},
  author={OpenAI},
  year={2025},
  url={https://api.semanticscholar.org/CorpusID:278283461}
}

@article{mcdermott1987student,
  title={Student difficulties in connecting graphs and physics: Examples from kinematics},
  author={McDermott, Lillian C and Rosenquist, Mark L and Van Zee, Emily H},
  journal={American journal of physics},
  volume={55},
  number={6},
  pages={503--513},
  year={1987},
  publisher={American Association of Physics Teachers}
}

@article{beichner1994testing,
  title={Testing student interpretation of kinematics graphs},
  author={Beichner, Robert J},
  journal={American journal of Physics},
  volume={62},
  number={8},
  pages={750--762},
  year={1994},
  publisher={[Woodbury, NY, etc. Published for the American Association of Physics~…}
}

@article{van1991learning,
  title={Learning to think like a physicist: A review of research-based instructional strategies},
  author={Van Heuvelen, Alan},
  journal={American Journal of physics},
  volume={59},
  number={10},
  pages={891--897},
  year={1991},
  publisher={American Association of Physics Teachers}
}

@article{english1997development,
  title={The development of fifth-grade children's problem-posing abilities},
  author={English, Lyn D},
  journal={Educational studies in Mathematics},
  volume={34},
  number={3},
  pages={183--217},
  year={1997},
  publisher={Springer}
}

@article{silver1994mathematical,
  title={On mathematical problem posing},
  author={Silver, Edward A},
  journal={For the learning of mathematics},
  volume={14},
  number={1},
  pages={19--28},
  year={1994},
  publisher={JSTOR}
}

@article{stoyanova1996framework,
  title={A framework for research into students’ problem posing in school mathematics},
  author={Stoyanova, Elena and Ellerton, Nerida F},
  journal={Technology in mathematics education},
  volume={4},
  number={7},
  pages={518--525},
  year={1996}
}

@article{singer2013problem,
  title={Problem-posing research in mathematics education: New questions and directions},
  author={Singer, Florence Mihaela and Ellerton, Nerida and Cai, Jinfa},
  journal={Educational studies in mathematics},
  volume={83},
  number={1},
  pages={1--7},
  year={2013},
  publisher={Springer}
}

@article{Polverini_2025,
   title={Performance of ChatGPT on tasks involving physics visual representations: The case of the brief electricity and magnetism assessment},
   volume={21},
   ISSN={2469-9896},
   url={http://dx.doi.org/10.1103/PhysRevPhysEducRes.21.010154},
   DOI={10.1103/physrevphyseducres.21.010154},
   number={1},
   journal={Physical Review Physics Education Research},
   publisher={American Physical Society (APS)},
   author={Polverini, Giulia and Melin, Jakob and Önerud, Elias and Gregorcic, Bor},
   year={2025},
   month=May }

@article{Dahlkemper_2023,
   title={How do physics students evaluate artificial intelligence responses on comprehension questions? A study on the perceived scientific accuracy and linguistic quality of ChatGPT},
   volume={19},
   ISSN={2469-9896},
   url={http://dx.doi.org/10.1103/PhysRevPhysEducRes.19.010142},
   DOI={10.1103/physrevphyseducres.19.010142},
   number={1},
   journal={Physical Review Physics Education Research},
   publisher={American Physical Society (APS)},
   author={Dahlkemper, Merten Nikolay and Lahme, Simon Zacharias and Klein, Pascal},
   year={2023},
   month=June }

@article{de2026stem,
  title={STEM Faculty Perspectives on Generative AI in Higher Education},
  author={de Silva, Akila and Song, Isabel Hyo Jung and Yang, Hui and Humayoun, Shah Rukh},
  journal={arXiv preprint arXiv:2603.04001},
  year={2026}
}

@inproceedings{wang2024examining,
  title={Examining the potential and pitfalls of ChatGPT in science and engineering problem-solving},
  author={Wang, Karen D and Burkholder, Eric and Wieman, Carl and Salehi, Shima and Haber, Nick},
  booktitle={Frontiers in Education},
  volume={8},
  pages={1330486},
  year={2024},
  organization={Frontiers Media SA}
}

@article{bitzenbauer2023chatgpt,
  title={ChatGPT in physics education: A pilot study on easy-to-implement activities},
  author={Bitzenbauer, Philipp},
  journal={Contemporary Educational Technology},
  volume={15},
  number={3},
  pages={ep430},
  year={2023},
  publisher={Bastas}
}

@article{ding2023students,
  title={Students’ perceptions of using ChatGPT in a physics class as a virtual tutor},
  author={Ding, Lu and Li, Tong and Jiang, Shiyan and Gapud, Albert},
  journal={International Journal of Educational Technology in Higher Education},
  volume={20},
  number={1},
  pages={63},
  year={2023},
  publisher={Springer}
}

@article{horchani2025chatgpt,
  title={ChatGPT’s problem-solving abilities in context-rich and traditional physics problems},
  author={Horchani, Ridha},
  journal={Physics Education},
  volume={60},
  number={2},
  pages={025019},
  year={2025},
  publisher={IOP Publishing}
}

@article{PhysRevPhysEducRes.19.010132,
  title = {Could an artificial-intelligence agent pass an introductory physics course?},
  author = {Kortemeyer, Gerd},
  journal = {Phys. Rev. Phys. Educ. Res.},
  volume = {19},
  issue = {1},
  pages = {010132},
  numpages = {18},
  year = {2023},
  month = {May},
  publisher = {American Physical Society},
  doi = {10.1103/PhysRevPhysEducRes.19.010132},
  url = {https://link.aps.org/doi/10.1103/PhysRevPhysEducRes.19.010132}
}

@article{nokes2013toward,
  title={Toward a model of transfer as sense-making},
  author={Nokes-Malach, Timothy J and Mestre, Jose P},
  journal={Educational Psychologist},
  volume={48},
  number={3},
  pages={184--207},
  year={2013},
  publisher={Taylor \& Francis}
}

@article{bransford1999chapter,
  title={Chapter 3: Rethinking transfer: A simple proposal with multiple implications},
  author={Bransford, John D and Schwartz, Daniel L},
  journal={Review of research in education},
  volume={24},
  number={1},
  pages={61--100},
  year={1999},
  publisher={Sage Publications Sage CA: Thousand Oaks, CA}
}

@article{chi1989self,
  title={Self-explanations: How students study and use examples in learning to solve problems},
  author={Chi, Michelene TH and Bassok, Miriam and Lewis, Matthew W and Reimann, Peter and Glaser, Robert},
  journal={Cognitive science},
  volume={13},
  number={2},
  pages={145--182},
  year={1989},
  publisher={Elsevier}
}

@article{booth2013using,
  title={Using example problems to improve student learning in algebra: Differentiating between correct and incorrect examples},
  author={Booth, Julie L and Lange, Karin E and Koedinger, Kenneth R and Newton, Kristie J},
  journal={Learning and Instruction},
  volume={25},
  pages={24--34},
  year={2013},
  publisher={Elsevier}
}

@book{miles2014qualitative,
  title={Qualitative Data Analysis: A Methods Sourcebook},
  author={Miles, Matthew B. and Huberman, A. Michael and Salda{\~n}a, Johnny},
  year={2014},
  edition={3},
  publisher={SAGE}
}

@article{de2021comparison,
  title={A comparison of reliability coefficients for ordinal rating scales},
  author={de Raadt, Alexandra and Warrens, Matthijs J and Bosker, Roel J and Kiers, Henk AL},
  journal={Journal of Classification},
  volume={38},
  number={3},
  pages={519--543},
  year={2021},
  publisher={Springer}
}

@article{shojaee2025illusion,
  title={The Illusion of Thinking: Understanding the Strengths and Limitations of Reasoning Models via the Lens of Problem Complexity},
  author={Shojaee, Parshin and Mirzadeh, Iman and Alizadeh, Keivan and Horton, Maxwell and Bengio, Samy and Farajtabar, Mehrdad},
  journal={arXiv preprint arXiv:2506.06941},
  year={2025}
}

@article{6fmx-bsnl,
  title = {Evaluating GPT- and reasoning-based large language models on Physics Olympiad problems: Surpassing human performance and implications for educational assessment},
  author = {Tschisgale, Paul and Maus, Holger and Kieser, Fabian and Kroehs, Ben and Petersen, Stefan and Wulff, Peter},
  journal = {Phys. Rev. Phys. Educ. Res.},
  volume = {21},
  issue = {2},
  pages = {020115},
  numpages = {21},
  year = {2025},
  month = {Aug},
  publisher = {American Physical Society},
  doi = {10.1103/6fmx-bsnl},
  url = {https://link.aps.org/doi/10.1103/6fmx-bsnl}
}

@article{casal2023ai,
  title={AI literacy in K-12: a systematic literature review},
  author={Casal-Otero, Lorena and Catala, Alejandro and Fern{\'a}ndez-Morante, Carmen and Taboada, Maria and Cebreiro, Beatriz and Barro, Sen{\'e}n},
  journal={International Journal of STEM Education},
  volume={10},
  number={1},
  pages={29},
  year={2023},
  publisher={Springer}
}

@inproceedings{Bralin2025, 
    Author = "Amir Bralin and N. Sanjay Rebello",
    Title = {AI Reasoning Models for Problem Solving in Physics}, 
    BookTitle = {Physics Education Research Conference 2025},
    Pages = {99-104},
    Address = {Washington, DC},
    Series = {PER Conference},
    Month = {August 6-7},
    Year = {2025}
}

\end{document}